\documentclass[journal]{IEEEtran} 

\usepackage{graphicx,epsfig,color,bm}
\usepackage{color,cite}
\usepackage{amsmath,epsfig} % easier math formulae, align, subequation
\usepackage{times}
\usepackage{enumerate,type1cm}
\usepackage{amsfonts,relsize}
\usepackage{bm}
\usepackage{amssymb}
\usepackage{relsize}
\usepackage{fancybox}
\usepackage{algpseudocode}
\usepackage[english]{babel}
\def\BibTeX{{\rm B\kern-.05em{\sc i\kern-.025em b}\kern-.08em
    T\kern-.1667em\lower.7ex\hbox{E}\kern-.125emX}}

\long\def\symbolfootnote[#1]#2{\begingroup%
\def\thefootnote{\fnsymbol{footnote}}\footnote[#1]{#2}\endgroup}

\newcommand{\beq}{\begin{equation}}
\newcommand{\eeq}{\end{equation}}
\newcommand{\beqa}{\begin{eqnarray}}
\newcommand{\eeqa}{\end{eqnarray}}

\newcommand{\abs}[1]{\left\vert#1\right\vert}
\newcommand{\norm}[1]{\Vert#1\Vert}

\newcommand{\pr}{{p}}

\newcommand{\veca}{\mathbf{a}}
\newcommand{\vecb}{\mathbf{b}}
\newcommand{\vecu}{\mathbf{u}}

\newcommand{\vecn}{\mathbf{n}}

\newcommand{\vecy}{\mathbf{y}}

\newcommand{\vecx}{\mathbf{x}}

\newcommand{\bmSigma}{\bm{\Sigma}}

\newcommand{\matA}{\mathbf{A}}

\newcommand{\matX}{\mathbf{X}}
\newcommand{\matY}{\mathbf{Y}}

\newcommand{\matP}{\mathbf{P}}

\newcommand{\matM}{\mathbf{M}}
\newcommand{\matN}{\mathbf{N}}
\newcommand{\matG}{\mathbf{G}}
\newcommand{\matJ}{\mathbf{J}}
\newcommand{\matI}{\mathbf{I}}

\newcommand{\Ex}{\mathbb{E}}

\newcommand{\bmgm}{\bm{\gamma}}
\newcommand{\bmGm}{\bm{\Gamma}}

\newcommand{\bmPhi}{\bm{\Phi}}
\newcommand{\bmLambda}{\bm{\Lambda}}
\newcommand{\bmmu}{\bm{\mu}}

\newcommand{\lb}{\left(}
\newcommand{\rb}{\right)}

\usepackage{float}
\usepackage{colortbl}

\DeclareMathOperator*{\argmin}{arg\,min}
\DeclareMathOperator*{\argmax}{arg\,max}

\title{Unlabelled Sensing: A Sparse Bayesian Learning Approach}
\author{
\authorblockN{Ranjitha Prasad}\\
\authorblockA{\begin{tabular}{cc}
TCS Innovation Labs, Gurgoan, India\\
\{ranjitha.prasad\}@tcs.com\\
\end{tabular}}
\vspace{-7mm}
}
\begin{document}

\maketitle
\thispagestyle{empty}
\begin{abstract}
We address the recovery of sparse vectors in an overcomplete, linear and noisy multiple measurement framework, where the measurement matrix is known upto a permutation of its rows. We derive sparse Bayesian learning (SBL) based updates for joint recovery of the unknown sparse vectors and the sensing order, represented using a permutation matrix. We model the sparse vectors using multiple  uncorrelated and correlated vectors, and in particular, we use the first order AR model for the correlated sparse vectors. We propose the Permutation-MSBL and a  Kalman filtering based Permutation-KSBL algorithm for low-complexity joint recovery of the sparse vectors and  the permutation matrix. The novelty of this work is in providing a simple update step for the permutation matrix using the rearrangement inequality. We demonstrate the mean square error and the permutation recovery performance of the proposed algorithms  as compared to a compressed sensing based scheme.
\end{abstract}

\textbf{EDICS}: SAS-STAT, MLSAS-SPARSE, MLSAS-BAYES, SAS-ADAP
\vspace{-4mm}
\section{Introduction and System Model}
Efficient techniques for reconstructing sparse signals from an  overcomplete system of linear equations using compressed sensing and Bayesian methods have received considerable attention in recent years \cite{Donoho, Tipping_RVM, BCS}. In a general multiple measurement setting, the observation matrix $\matY \in \mathbb{R}^{N \times M}$ is obtained as a weighted combination of the columns of the measurement matrix $\bmPhi$, where the weights are given by the entries of $M$ sparse columns in $\matX \in \mathbb{R}^{L \times M}$ \cite{wipfRao_MMV}, i.e.,
\begin{equation}
\matY = \bmPhi\matX + \matN,
\vspace{-2mm}
\label{signal_modelconv}
\end{equation}
where each column $\vecn_m$ of the additive noise matrix $\matN \in \mathbb{R}^{N \times M}$ is modelled as a zero mean additive white Gaussian vector distributed as $\vecn_m \sim \mathcal{N}(0,\sigma^2\matI_N)$. In \eqref{signal_modelconv}, the columns of $\matX$ have identical sparsity profiles, i.e., $\matX$ has $K$ non-zero rows, where $K$ is the \emph{sparsity} or the number of non-zero entries in each column of $\matX$.

Typically, the task of sparse recovery involves the recovery of the locations and the magnitude of the non-zero entries of the sparse vector when  the observation matrix is perfectly known. In the case when the observation matrix is unknown or partially known, several papers in literature have addressed the problem of jointly recovering the observation matrix and the sparse vectors \cite{prasadjointSISO, parkerVolcanSchniter, prasadjointMIMO,sadeghidictionary}. In this work, we address a special case of a partially known observation matrix, wherein the true order, or the \emph{sensor permutation} of rows of the observation matrix is unknown. This problem is called as the \emph{sensor permutation} (SP) problem or unlabelled sensing \cite{Wainwright,Unnikrish,Gibronv}. 

Unlabelled sensing problems are encountered in several  applications such as simultaneous location approximation and mapping in robotics, multi-target tracking, archaeological measurements, clock jitter  \cite{Wainwright, Caire2017}, and more recently in device-to-device communications \cite{wieruchD2D}. Sparse linear unlabelled sensing problem using compressed sensing based techniques was explored recently. In \cite{tillmanncomputational}, the authors consider the problem of joint SP dictionary and  sparse vector estimation and discuss several results pertaining to its computational intractibility. In \cite{Unnikrish}, the authors address the identifiability of $\matX$ in a \emph{noiseless} setting, and show that when the number of observations is twice the sparsity level, exact recovery is possible with probability $1$. In \cite{Wainwright}, the authors provide sharp conditions on the signal to noise ratio (SNR) for exact permutation recovery, and also derive necessary conditions for approximate permutation recovery when the entries of the observation matrix are drawn from a standard Gaussian matrix in the \emph{noisy} linear setting. For unlabeled sensing problems where the relative order of the subset of observations is known, an alternating maximization algorithm is proposed in  \cite{Caire2017}. A branch and bound based solution to the noiseless MMV SP problem was proposed in \cite{Gibronv}. In \cite{abid2017linreg},  the authors propose a least squares and a methods of moments based estimator for recovering the true sensor permutation. In a slightly different SP problem setting, algorithms to recover the unknown sparse vector and parameters from binary quantized measurements are proposed  in \cite{wang2017}. However, none of the above mentioned works consider a Bayesian framework where the sparse matrix $\matX$ is drawn from a given distribution. 

We propose a novel Bayesian  approach for joint multiple sparse vector recovery and sensor permutation recovery in an MMV framework. In particular, we solve the  SP problem in an SBL framework \cite{Tipping_RVM, Wipf_Rao}, and derive the update equations for the unknown sparse vectors and the permutation matrices. The observations $\vecy_m \in \mathbb{R}^{N \times 1}$ is linearly related to the unknown \emph{permutation matrix} $\matP_m$ and the sparse vectors in $\matX$ as
\begin{align}
&\vecy_{m} = \matP_m\bmPhi\vecx_m+\vecn_{m}, \nonumber\\
&\vecx_{m+1} = \rho \vecx_{m} + \vecu_{m+1}, \quad m = 1, 2, \hdots, M.
\label{system_statemeasumodel}
\end{align}
Here $\matP_m \in \mathcal{P}_N$, where $\mathcal{P}_N$ is the set of all $N \times N$ permutation matrices. The prior in the context of SBL is given by $p(\matX; \bmGm) = \prod_{m = 1}^M p(\vecx_m;\bmGm)$, where  $\bmGm = \mbox{diag}(\gamma(1),\hdots,\gamma(L))$ and 
\begin{equation}
\pr(\vecx_m; \bmGm) = \prod_{l = 1}^{L} (2 \pi \gamma(l))^{-\frac{1}{2}} \exp\lb- \frac{\abs{x_m(l)}^2}{2 \gamma(l)}\rb.
\label{Gaussian_prior}
\end{equation}
Note that if $\gamma(l) \rightarrow 0$, then the corresponding $x(l) \rightarrow 0$~\cite{Tipping_RVM, Wipf_Rao}, and $\bmgm$ is $K$-sparse.
In \eqref{system_statemeasumodel}, we employ a first order auto-regressive (AR) model for modelling the correlation between adjacent columns, $\vecx_m$ and $\vecx_{m+1}$ in $\matX$, where $0 \leq \rho \leq 1$ is the AR coefficient and $\vecu_m$ is the driving noise which is typically modeled as $u_m \sim \mathcal{N}(0,(1-\rho^2)\Gamma)$. The columns of $\matX$ in \eqref{system_statemeasumodel} are statistically independent of each other if $\rho = 0$.  In the model given above, the sparsity of driving noise $\vecu_m$ in \eqref{system_statemeasumodel} is coupled to the sparsity in $\bmgm$, i.e., if $\gamma(l) \rightarrow 0$, then the corresponding $u_m(l) \rightarrow 0$. 

In the sequel, we propose SBL based algorithms as a solution to the SP problem and demonstrate the performance using Monte carlo simulations. To the best of our knowledge, this is the first work that addresses the SP problem in a  Bayesian framework. 

\textbf{Notation:} Boldface small letters denote vectors and boldface capital letters denote matrices. The symbol $\mbox{diag}(\veca)$ denotes a diagonal matrix with entries given by $\veca$. The pdf of the random variable $X$ is represented as $\pr(x)$ and the random variables and deterministic parameters in the pdf are separated using a semicolon. The $i^{\text{th}}$ entry of $\veca$ and the $(i,j)^{\text{th}}$ entry of $\matA$ are represented as $a(i)$ and $A(i,j)$, respectively.
%In the sequel, we obtain SBL based algorithms for the SP problem in an MMV framework, and we show that the optimization problem for the permutation matrix is easily solved by invoking the rearrangement inequality. \\
%\textbf{Notation:}

%Using small number of anchor rows of the permutation matrix, we obtain reliable estimates of hyperparameters and learn the remaining entries of the permutation matrix. 
%%Typically, the hyperparameters $\bmgm$ can be estimated using a type-II ML procedure \cite{wipfRao_MMV}, i.e., by maximizing the marginalized pdf $\pr(\matY;\bmGm)$.
%\begin{equation}
%\hat{\bmgm}_{ML} = \argmax_{\bmgm \in \mathbb{R}^L_{+}} \pr(\matY;\bmGm).
%\end{equation}

%\section{Proposed Approach}
%In this section, we propose two novel SBL based solutions for joint recovery of the permutation matrix $\matP$, and the multiple sparse vectors $\matX = [\vecx_1, \hdots, \vecx_M]$, where $\vecx_i$ is the $i$-th sparse column of $\matX$. 
 
%Often, it is not straightforward to solve the above problem in closed form and hence, iterative estimators such as the expectation maximization (EM) based SBL algorithm \cite{Wipf_Rao} are employed. 

\section{P-MSBL: Uncorrelated Sparse Columns of $\matX$}
\label{PMSBL}
We propose the \emph{\textbf{P}ermutation}-MSBL (P-MSBL) algorithm for joint recovery of the unknown permutation matrix $\matP$ and multiple sparse vectors in $\matX$, when $\rho = 0$. We treat $\matX$ as the hidden variable, and $(\matY,\matX)$ as the complete information, and both $\bmgm$ and $\matP = [\matP_1,\hdots,\matP_M]$ are unknown parameters. We obtain the maximum likelihood estimate of the unknown parameters using the SBL algorithm as follows:
\begin{align}
&\mbox{E}:Q\left(\bmgm, \matP|\bmgm^{(r)}, \matP^{(r)}\right) = \Ex_{\matX|\matY;\bmgm^{(r)}, \matP^{(r)}}[\log \pr(\matY,\matX;\bmgm,\matP)]\nonumber\\
&\mbox{M}: \bmgm^{(r+1)},\matP^{(r+1)} = \argmax_{\bmgm \in \mathbb{R}^{L \times 1}_{+}, \matP \in \mathcal{P}_N^M} Q\left(\bmgm,\matP|\bmgm^{(r)},\matP^{(r)}\right),
\label{EM_unksenperm}
\end{align}
where $\mathcal{P}^M_N$ is the $M$ dimensional manifold of $\mathcal{P}_N$. The E-step above requires the posterior density of the sparse vector using the $r$-th parameter updates given by $\bmgm^{(r)}$ and $\matP^{(r)}$. Using the likelihood given by
\begin{equation}
p(\vecy_m|\vecx_m;\matP_m) = \frac{1}{(2 \pi \sigma^2)^{M/2}}\exp\left(-\frac{\norm{\vecy_m - \matP_m\bmPhi\vecx_m}^2}{2\sigma^2}\right),
\end{equation}
and the prior density in \eqref{Gaussian_prior}, the posterior distribution is computed as $\pr\left(\vecx_m|\vecy_m; \bmgm^{(r)}, \matP_m^{(r)}\right) \sim \mathcal{N}(\bmmu_m,\bmSigma_m)$, where
\begin{align}
\bmSigma_m &= \bmGm^{(r)} -\bmGm^{(r)}\bmPhi^T ({\matP_m^{(r)}})^T \bmLambda_m^{-1}\matP_m^{(r)}\bmPhi\bmGm^{(r)}\nonumber\\
\bmmu_m &= \sigma^{-2}\bmSigma_m \bmPhi^T ({\matP_m^{(r)}})^T \vecy_m, ~~ m \in [M],
\label{muSigmaPMSBL}
\end{align}
where $\bmLambda_m = \left(\sigma^2 \mathbf{I}_{N} + \matP_m^{(r)}\bmPhi  \bmGm^{(r)} \bmPhi^T ({\matP_m^{(r)}})^T \right)$. In the above equations, $\bmmu_m$ is the MAP estimate of $\vecx_m$ for the $r$-th update of the  hyperparameters given by $\bmGm^{(r)} = \mbox{diag}(\bmgm^{(r)})$. In the noiseless case the SBL updates can be derived by applying the limit $\sigma^2 \rightarrow 0$ \cite{wipfRao_MMV}.

%The M-step in \eqref{EM_unksenperm} can be simplified to obtain
%\begin{align} 
%\bmgm^{(r+1)}, \matP^{(r+1)} &= \argmax_{\bmgm \in \mathcal{R}^{L \times 1}_+, \matP \in \mathcal{P}} \Ex_{\matX|\matY;\bmgm^{(r)}, \matP^{r}}\left[\log \pr(\matY,\matX;\bmgm,\matP)\right]
%\label{MLprob_gamma}
%\end{align}
The key aspect of the M-step given in \eqref{EM_unksenperm} is that the objective function is the sum of two independent functions
\begin{equation}
Q\left(\bmgm, \matP_m|\bmgm^{(r)}, \matP_m^{(r)}\right) = Q(\bmgm| \bmgm^{(r)}) + \sum_{m = 1}^MQ(\matP_m|\matP_m^{(r)}),
\end{equation}
where $Q(\matP_m|\matP_m^{(r)}) = \Ex_{\matX|\matY;\bmgm^{(r)},\matP_m^{(r)}}[\log \pr(\matY|\matX;\matP_m)]$ and $Q(\bmgm| \bmgm^{(r)}) = \Ex_{\matX|\matY;\bmgm^{(r)},\matP_m^{(r)}}[\log\pr(\matX;\bmgm)]$, hence admitting independent optimization w.r.t. $\bmgm$ and $\matP_m$ for $m \in [M]$. \\
\textbf{Optimization w.r.t. $\matP_m$}: The optimization problem to obtain the permutation matrix update, $\matP_m^{(r+1)}$, is as follows
\begin{align}
&\matP_m^{(r+1)} = \argmax_{\matP_m \in \mathcal{P}_N}~\Ex_{\matX|\matY;\bmgm^{(r)},\matP^{(r)}}[\log \pr(\matY|\matX;\matP_m)] \\
& = \argmin_{\matP_m \in \mathcal{P}_N}~\Ex_{\matX|\matY;\bmgm^{(r)},\matP^{(r)}}[\norm{\vecy_m - \matP_m\bmPhi\vecx_m}_2^2]\\
& = \argmin_{\matP_m \in \mathcal{P}_N}~c - \Ex_{\matX|\matY;\bmgm^{(r)},\matP^{(r)}}[ \vecy_m^T\matP_m\bmPhi\vecx_m]\label{Pmatlaststep},
\end{align}
where $c$ consists of terms independent of $\matP_m$ for $m \in [M]$. We obtain \eqref{Pmatlaststep} using $\matP_m^T\matP_m = \matI_N$. From the E-step, we have $\Ex_{\matX|\matY;\bmgm^{(r)},\matP^{(r)}}[\vecx_m] = \bmmu_m$, and hence, \eqref{Pmatlaststep} is re-written as
\begin{align}
\matP_m^{(r+1)} = \argmax_{\matP_m \in \mathcal{P}_N} ~~ \vecy_m^T\matP_m\bmPhi\bmmu_m.
\label{Pmatoptiprob}
\end{align}

In summary, the solution to the above optimization problem maximizes  the $m$-th term which is the product of two vectors $\vecy_m$ and $\bmPhi\bmmu_m$, i.e., we need to find the permutation matrix $\matP_m$ which  permutes the rows of $\bmPhi\bmmu_m$ (or columns of $\vecy_m^T$) such that the product of two vectors is maximized.  The solution for  \eqref{Pmatoptiprob} can be obtained by invoking the \emph{rearrangement inequality}\cite{hardypolyainequalities, Wainwright}. This  inequality states that if $\veca = (a_1, a_2, \hdots, a_n)$ and $\vecb = (b_1, b_2, \hdots , b_n)$ are two $n$-tuples of real numbers, and $\sigma_{\veca}$ and $\sigma_{\vecb}$ are permutations on $\veca$ and $\vecb$ such that $\sigma_{\veca}(a_1)\geq \hdots \geq \sigma_{\veca}(a_n)$ and $\sigma_{\vecb}(b_1) \geq \hdots \geq \sigma_{\vecb}(b_n)$, and $\sigma'_{\veca}$ and $\sigma'_{\vecb}$ be arbitrary permutations of natural numbers on $\veca$ and $\vecb$, then
\begin{equation}
\sum_i \sigma_\veca(a_i) \sigma_\vecb(b_i) \geq \sum_i \sigma'_\veca(a_i) \sigma'_\vecb(b_i).
\end{equation}
In other words, for two vectors $\veca$ and $\vecb$, the sum of pairwise products is maximum if we pair the largest element of the first vector with the largest element of the second vector, the second largest element of the first vector with the second largest element of the second vector, and so on. 

Invoking the rearrangement inequality in this context, the permutation matrix for every $m$ is obtained by sorting the vectors $\vecy_m$ and $\bmPhi\bmmu_m$ in descending (or ascending) order, leading to $M$ permutation matrices. If $\matP_1 = \hdots = \matP_M = \bar{\matP}$, we obtain the following optimization problem in $\matP$:
\begin{align}
\bar{\matP}^{(r+1)} = \argmax_{\bar{\matP} \in \mathcal{P}_N} ~~ \sum_{m = 1}^M\vecy_m^T\bar{\matP}\bmPhi\bmmu_m.
\label{PmatoptiprobsameP}
\end{align}
It is possible to solve the above problem by obtaining a convex relaxation of \eqref{PmatoptiprobsameP} by replacing $\mathcal{P}_N$ by its convex hull. However, it is shown that in practice, this strategy performs poorly and fails to identify the true permutation \cite{blumesmathDavies, Gibronv}. For the EM algorithm to converge, it is sufficient to ensure that the likelihood increases in every iteration of the algorithm, i.e.,
\begin{align}
Q\left(\bmgm^{(r+1)}, \bar{\matP}^{(r+1)}|\bmgm^{(r)}, \bar{\matP}^{(r)}\right) \geq Q\left(\bmgm, \bar{\matP}|\bmgm^{(r)}, \bar{\matP}^{(r)}\right).
\end{align}
In order to ensure a likelihood increase, we propose a low-complexity albeit an approximate solution to \eqref{Pmatoptiprob} by choosing $M$ candidate permutation matrices that maximize each of the $M$ terms in \eqref{PmatoptiprobsameP}. Among the $M$ candidate solutions, we choose the permutation matrix that leads to the largest value of \eqref{PmatoptiprobsameP}, and hence the likelihood. \\
% as follows:
%\begin{equation}
%\matP(\sigma_{\vecy_m}(\vecy_m(i)),\sigma_{(\bmPhi\bmmu_m)}((\bmPhi\bmmu_m)(i)) = 1,
%\end{equation}
%for $1 \leq i \leq N$, and all other entries of $\matP$ are set to $0$. Here $(\bmPhi\bmmu_m)(i)$ represents the $i$-th term of $\bmPhi\bmmu_m$. \\
\textbf{Optimization w.r.t. $\bmgm$}: The optimization problem w.r.t. $\bmgm$ is given by
\begin{equation}
\bmgm^{(r+1)} = \argmax_{\bmgm \in \mathcal{R}^{L \times 1}_+}~~ \Ex_{\matX|\matY;\bmgm^{(r)},\matP^{(r)}}[\log\pr(\matX;\bmgm)].
\label{gammaOpti}
\end{equation}
Optimizing \eqref{gammaOpti} w.r.t. $\gamma(l) \in [L]$, we obtain the following:
\begin{align}
\gamma^{(r+1)}(l) = \frac{1}{M}\sum_{m = 1}^M\left(\bmSigma_m(l,l) +\norm{\bmmu_m(l)}_2^2\right),
\end{align}
where $\bmSigma_m(l,l)$ and $\bmmu_m(l)$ are the $l$-th diagonal component and the $l$-th component of $\bmSigma_m$ and $\bmmu_m$, respectively. 

In the following section, we derive the Kalman filtering based SBL-based updates for the system model given in \eqref{system_statemeasumodel}.
%\vspace{-3mm}

\section{P-KSBL: Correlated Sparse Columns of $\matX$}

We design SBL based algorithms for the joint recovery of the correlated sparse vectors and the permutation matrices. Specifically, we consider the scenario when $0 < \rho \leq 1$, and provide SBL updates which exploit the correlation between the columns of $\matX$. Note that for $\rho = 0$, the Kalman filter based update equations are same as in \eqref{muSigmaPMSBL}. 
The E- and the M-steps of the SBL algorithm are as given in \eqref{EM_unksenperm}. If the measurement matrices and the hyperparameters are known, the model in \eqref{system_statemeasumodel} admits a Kalman filter based estimation and tracking framework. This motivates us to design such a filter using SBL updates. Accordingly, we observe that in the E-step given in \eqref{EM_unksenperm}, posterior distribution of the sparse vector, $p(\matX|\matY; \bmgm^{(r)}, \matP^{(r)})$, is obtained using the recursive update equations given by  \cite{prasadjointSISO}
\begin{align}
&\mbox{\textbf{for}}~m = 1,\hdots,M ~~ \mbox{\textbf{do}}\nonumber\\
&\mbox{Prediction:} ~ \hat{\vecx}_{m|m-1} = \rho \hat{\vecx}_{m-1|m-1}\label{kalman_update_start}\\ 
&\bmSigma_{m|m-1} = \rho^2 \bmSigma_{m-1|m-1} + (1-\rho^2)\bmGm^{(r)} \label{pred_var}\\
&\mbox{Filtering:} ~\matG_m = \bmSigma_{m|m-1}\bmPhi_{m}^T\nonumber\\
&~~~(\matP_m^{(r)})^T\left(\sigma^2 \mathbf{I}_{N} + \matP_m^{(r)}\bmPhi_{m}\bmSigma_{m|m-1}\bmPhi_{m}^T(\matP_m^{(r)})^T\right)^{-1} \label{Kalman_gain}\\
&\hat{\vecx}_{m|m} =\hat{\vecx}_{m|m-1} + \matG_m(\vecy_{m} - \matP_m^{(r)}\bmPhi_{m} \hat{\vecx}_{m|m-1})\label{chan_update}\\
&\bmSigma_{m|m} = (\mathbf{I}_{L} - \matG_m\matP_m^{(r)}\bmPhi_{m})\bmSigma_{m|m-1}~~~\mbox{\textbf{end}}\label{kalman_update_end} \\ 
&\mbox{\textbf{for}}~j = M, M-1, \hdots, 2~~ \mbox{\textbf{do}}\nonumber\\
&\mbox{Smoothing:} ~ \hat{\vecx}_{j-1|M} = \hat{\vecx}_{j-1|j-1} + \matJ_{j-1}( \hat{\vecx}_{j|M} -  \hat{\vecx}_{j|j-1})\label{kalman_smooth_start}\\
&\bmSigma_{j-1|M} = \bmSigma_{j-1|j-1} + \matJ_{j-1}(\bmSigma_{j|M} - \bmSigma_{j|j-1})\matJ_{j-1}^T \label{kalman_smooth_end}\\
&\mbox{\textbf{end}}. \nonumber 
\end{align}
Here $\matJ_{j-1} \triangleq \rho \bmSigma_{j-1|j-1}\bmSigma_{j|j-1}^{-1}$ and $\matG_m$ is the Kalman gain, and the symbols $\hat{\vecx}_{m|m-1}$, $\bmSigma_{m|m-1}$, etc.\ have their usual meanings as in the KF literature \cite{Ghahramani_mainlearning}. The above  procedure is initialized by setting  $\hat{\vecx}_{0|0} = \mathbf{0}$ and $\bmSigma_{0|0} = \bmGm^{(0)}$.

In order to obtain an ML estimate of $\bmgm$ and the permutation matrix $\matP_m$ in the M-step, the mean and covariance of the posterior distribution computed in the E-step is utilized. From \eqref{EM_unksenperm}, the M-step results in the following:
\begin{eqnarray}
&Q(\bmgm, \matP|\bmgm^{(r)}, \matP^{(r)}) =\nonumber\\
&\Ex_{\matX|\matY;\bmgm^{(r)},\matP^{(r)}}[c' - \sum\limits_{j = 1}^{M}\frac{\norm{\vecy_{j} - \matP_{j}\bmPhi_j\vecx_j}^2}{\sigma^2}- M\log |\Gamma | \nonumber\\
&- \sum\limits_{j = 2}^{M} \frac{(\vecx_j - \rho \vecx_{j-1})^T\Gamma^{-1}(\vecx_j - \rho \vecx_{j-1})}{(1-\rho^2)}- \vecx_1^T\Gamma^{-1}\vecx_1],
\end{eqnarray}
where $c'$ is a constant independent of $\bmgm$ and $\matP$. The expression above is a sum of terms which are independent functions of $\bmgm$ and $\matP$, denoted as $Q(\bmgm|\bmgm^{(r)}_{m})$ and $Q(\matP|\matP^{(r)})$, respectively. The update of the hyperparameters for the $(r+1)$-th iteration can be computed from the optimization problem given by 
\begin{align}
&\bmgm^{(r+1)} = \argmax_{\bmgm \in \mathbb{R}^{L \times 1}_{+}}\Ex_{\matX|\matY;\bmgm^{(r)}}[-M\log |\bmGm|\nonumber\\
& - \sum_{j = 2}^{M}  \frac{(\vecx_j - \rho\vecx_{j-1})^T\bmGm^{-1}(\vecx_j - \rho\vecx_{j-1})}{(1-\rho^2)} - \vecx_1^T\bmGm^{-1}\vecx_1].
\label{KSBL_mstep}
\end{align}
The above expression can be simplified as
\begin{align}
\gamma^{(r+1)}(l) = \frac{1}{M} (\sum_{j=2}^M\tfrac{\matM_{j|M}(l,l)}{(1-\rho^2)} + \matM_{1|M}(l,l)),
\label{hyper_update}
\end{align}
for $l \in [L]$, $\matM_{j|M} \triangleq \bmSigma_{j|M} + \hat{\vecx}_{j|M}\hat{\vecx}^T_{j|M}+\rho^2(\bmSigma_{j-1|M}+\hat{\vecx}_{j-1|M}\hat{\vecx}^T_{j-1|M}) - 2\rho\mbox{Re}(\bmSigma_{j,j-1|M}+\hat{\vecx}_{j|M}\hat{\vecx}^T_{j-1|M})$ and $\matM_{1|M} \triangleq \bmSigma_{1|M} + \hat{\vecx}_{1|M}\hat{\vecx}^T_{1|M}$. Here, $\hat{\vecx}_{j|M} \triangleq \Ex_{\matX|\matY;\bmgm^{(r)}}[\vecx_j]$, and  covariance $\Ex_{\matX|\matY;\bmgm^{(r)}}[\vecx_j\vecx_j^T] \triangleq \bmSigma_{j|M}+ \hat{\vecx}_{j|M}\hat{\vecx}^T_{j|M}$ for $j = 1, \hdots, M$ is obtained from \eqref{kalman_update_start}-\eqref{kalman_smooth_end}. Furthermore, $\Ex_{\matX|\matY;\bmgm^{(r)}}[\vecx_{j}\vecx_{j-1}^T] \triangleq \bmSigma_{j,j-1|M} + \hat{\vecx}_{j|M}\hat{\vecx}^T_{j-1|M}$ for $j = M, M-1,\hdots, 2$, which we obtain from \cite{Ghahramani_mainlearning} as follows:
\begin{eqnarray}
&\bmSigma_{j-1,j-2|M} = \bmSigma_{j-1|j-1}\matJ^T_{j-2} +\matJ^T_{j-1}(\bmSigma_{j,j-1|M}\nonumber\\
& - \rho \bmSigma_{j-1|j-1})\matJ_{j-2}.
\label{P_k-1k-2}
\end{eqnarray}
The above recursion is initialized using $\bmSigma_{M,M-1|M} = \rho (\mathbf{I}_{L} - \matG_M\matP_M\bmPhi_{M})\bmSigma_{M-1|M-1}$. 

The expression for $Q(\matP_m|\matP^{(r)}_m)$ for $m = 1, \hdots, M$ can be simplified as follows:
\begin{equation}
Q(\matP_m|\matP^{(r)}_m) =  \Ex_{\matX|\matY;\bmgm^{(r)}, \matP_m^{(r)}}[c - \frac{\norm{\vecy_m - \matP_m\bmPhi_m\vecx_m}^2}{\sigma^2}].
\label{Mstep_EMKSBL}
\end{equation}
From the above, we see that the M-step requires $\hat{\vecx}_{m|m} \triangleq \Ex_{\matX|\matY;\matP_m^{(r)},\bmgm^{(r)}}[\vecx_m]$ which is computed in the E-step. The maximization of  $Q(\matP_m|\matP^{(r)}_m)$ in \eqref{Mstep_EMKSBL} leads to the following optimization problem for~$\matP_m$:
\begin{align}
\matP_m^{(r+1)} = \argmax_{\matP_m \in \mathcal{P}_N} ~~ \vecy_m^T\matP_m\bmPhi_m\hat{\vecx}_{m|m}.
\label{Pmatoptiprobinm}
\end{align}
Again, we invoke the rearrangement inequality to solve the above optimization problem as detailed in the previous section. Thus, the proposed algorithm learns $\bmgm$ and $\matP_m$ in the M-step and provides low-complexity and recursive estimates of the sparse vector in the E-step. In the following section, we demonstrate the performance of the proposed algorithms using experimental results.

\section{Simulation Results}

In this section, we describe the experimental setup used to demonstrate the mean square error (MSE) and the permutation recovery performance (success rate) of the proposed algorithms. We consider $M$ sparse vectors of $L = 100$, with sparsity given by $K = 4$. We assume the perfect knowledge of the noise variance ($\sigma^2$) which we vary to capture a wide spectrum of signal to noise ratio. The matrix $\bmPhi$ is generated as an overcomplete ($N < L$) random Gaussian measurement matrix, with $N = 30$. In order to demonstrate the performance of the P-KSBL algorithm, we generate the correlated sparse vectors according to the AR model given in \eqref{system_statemeasumodel} with $\rho = 0.95$. The experiment is repeated for $1000$ trials. We fix the maximum number of iterations of the SBL algorithm as $500$. The algorithms proposed in this paper use EM-based updates, and hence, they have a local minima convergence guarantee~\cite{mclachlan_EMalgo}.

It is well-known that the SBL algorithm is particularly sensitive to the choice of the initialization parameters, $\bmGm^{(0)}$ and $\matP^{(0)}$. Using a  non-informative choice such as $\bmGm^{(0)} = \matI_L$ is sufficient for ensuring good solutions for $\bmgm$, when the measurement matrix is known \cite{prasadbayesianGlbm}. However, in the current scenario, the choice of $\matP^{(0)}$ is not very straightforward. In order to obtain $\matP^{(0)}$, we assume that $\matP$ consists of a few anchor rows which are known in advance. The anchor rows matrix $\matP_B$ consists of $B$ anchor rows of $\matP$ \cite{marques2009subspace}. Although the assumption on anchor rows is a departure from the derivation in Sec.~\ref{PMSBL}, an unfavourable choice of $\matP^{(0)}$ could have deflected the $\bmgm$ updates away from the true value. 
%First, we  obtain an intermediate estimate of $\bmgm$ using $\matP = \matP_B$. Subsequently, we obtain an initial estimate of $\matP^{(0)}$ using the intermediate estimate of $\bmgm$.

\begin{figure}[htp]
\begin{center}
\includegraphics[scale=0.4]{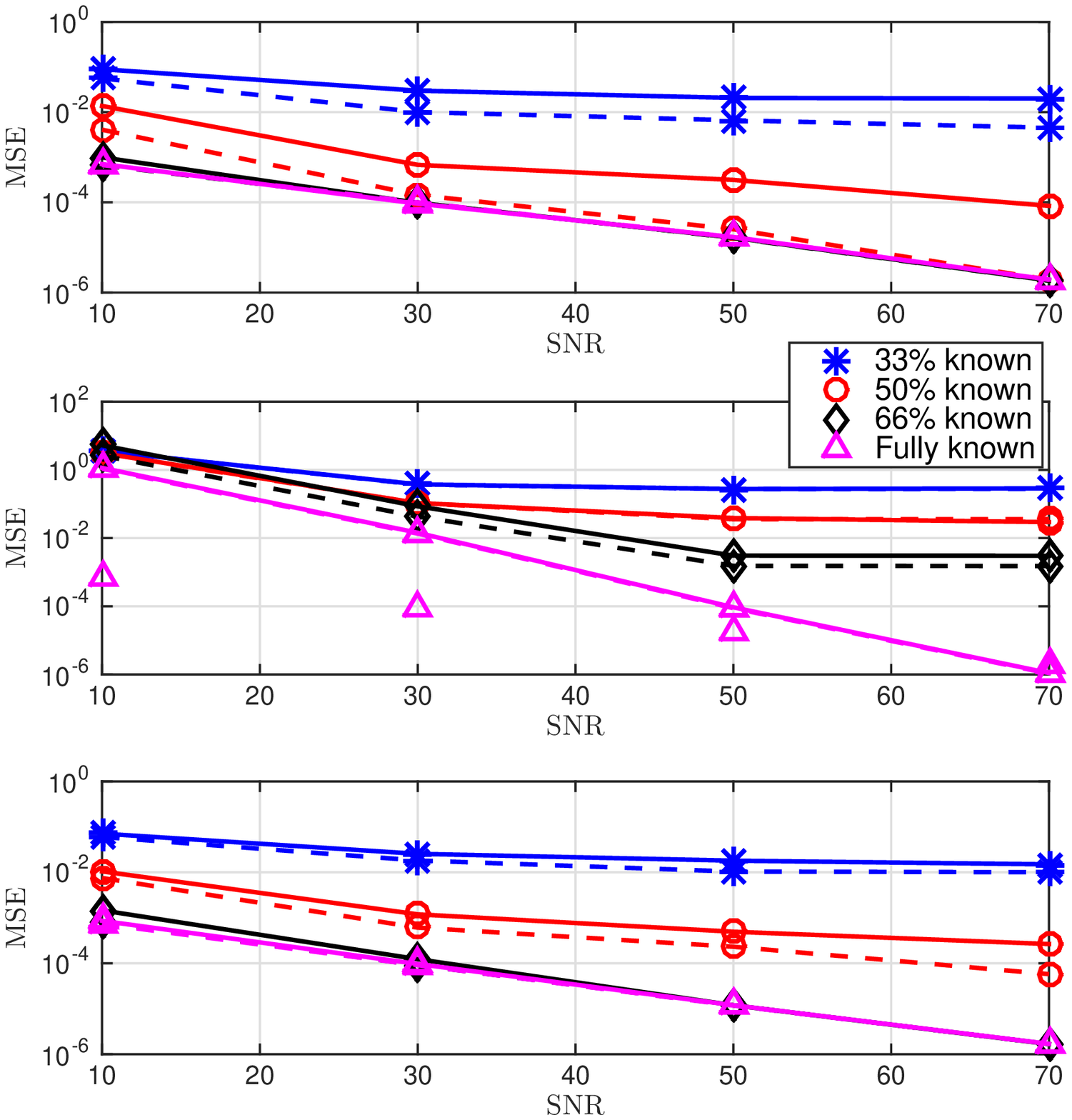}
\caption{Plot of the MSE performance of $\hat{\matX}(\matY)$ (Top:P-MSBL ($\matP_1 = \matP_2 = \hdots = \matP_M = \matP$), Middle: P-MSBL ($\matP_1 \neq \matP_2 \neq \hdots \neq \matP_M$), Bottom: P-KSBL ($\matP_1 = \matP_2 = \hdots = \matP_M = \matP$) for $N = 30$.}
\label{MSEMKSBL}
\end{center}
\end{figure}

In Fig.~\ref{MSEMKSBL}, we demonstrate the MSE performance of the proposed P-MSBL (when $\matP_1 \neq \matP_2 \neq \hdots \neq \matP_M$ and $\matP_1 = \matP_2 = \hdots = \matP_M = \matP$) and P-KSBL. Here, $\matP_m^{(0)}$ is as obtained above, $\bmGm^{(0)} = \matI_L$ and $B$ anchor rows of $\matP_m$ are assumed to be known apriori.  As expected, the performance of the proposed algorithms  improves as the number of anchor rows of $\matP$ increases. The performance also improves as $M$ increases. We also see that in the case when permutation matrices are unequal, the advantage of averaging over permutation matrices is not available, and hence increasing $M$ does not improve performance.  

\begin{figure}[htp]
\begin{center}
\includegraphics[scale=0.38]{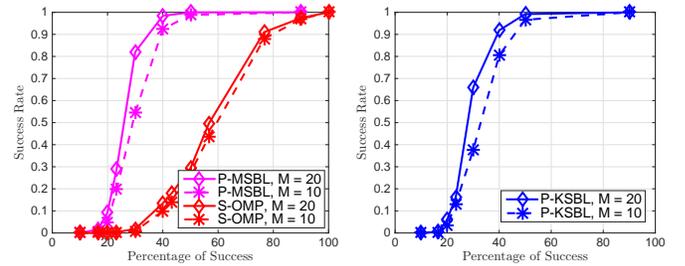}
\caption{Plot of the success rate of $\hat{\matX}(\matY)$ (Left:P-MSBL and S-OMP, Right: P-KSBL) for $SNR = 60$dB and varying values of $M$.}
\label{SuccessMKSBL}
\end{center}
\end{figure}

In Fig.~\ref{SuccessMKSBL}, the permutation recovery performance of the proposed techniques for $M = 20$ and $M = 10$ at SNR$ = 60$dB is demonstrated. Both, in the presence and absence of correlation, SBL can recover the permutations perfectly if $50\%$ of the permutation matrix consists of anchor rows, and the permutations are recovered with probability of $0.8$ if $30\%$ of the permutation matrix consists of anchor rows. It can be observed that the permutation recovery performance improves with $M$. As a baseline scheme, we employ the Simultaneous OMP (S-OMP) technique to obtain an estimate of $\matX$ using $\matP_B$. After obtaining the estimate of the sparse vector, the permutation matrix can be obtained by performing the M-step of the EM algorithm once. As the algorithm proceeds, $\bmgm$ and $\matP_m$ are estimated jointly and as a result, the accurately decoded rows of $\matP_m$ helps to improve the estimates of $\bmgm$. Hence, P-MSBL and P-KSBL perform better compared to  schemes that separately estimate $\matP_m$ and $\bmgm$, such as S-OMP.

%\begin{figure}[!htb]
%    \centering
%    \begin{minipage}{.3\textwidth}
%        \centering
%        \includegraphics[scale = 0.2]{SuccRateM10.eps}
%        \caption{$dt=0.1$}
%        \label{SuccRateM10}
%    \end{minipage}%
%    \begin{minipage}{0.3\textwidth}
%        \centering
%        \includegraphics[scale = 0.2]{SuccRateM20.eps}
%        \caption{$dt =$}
%        \label{SuccRateM20}
%    \end{minipage}
%\end{figure}
 
\vspace{-6mm}
\section{Conclusions}

We proposed novel SBL-based algorithms for recovery of a sparse matrix from a noisy, linear overcomplete MMV model, where the measurement matrix is known upto a permutation of rows. In particular, we considered two scenarios: a matrix with uncorrelated sparse columns and correlated sparse columns. We modelled the correlated sparse vectors using a first order AR model and devised a Kalman based SBL approach for joint recovery of the sparse vector and the unknown permutation matrix. An important aspect of the proposed algorithms is that the joint optimization problem in the hyperparameters of the sparse vector and the permutation matrix separates as two independent optimization problems. Furthermore, we simplified the permutation recovery problem by invoking the rearrangement inequality. Using Monte Carlo simulations, we showed that proposed techniques  are capable of learning the permutation matrix using a small number of known anchor rows in $\matP$.
\label{sec:concl}
\bibliographystyle{IEEEtran}
\bibliography{unknownPermbib}

\begin{thebibliography}{10}
\providecommand{\url}[1]{#1}
\def\UrlFont{\ttfamily}
\providecommand{\newblock}{\relax}
\providecommand{\bibinfo}[2]{#2}
\providecommand\BIBentrySTDinterwordspacing{\spaceskip=0pt\relax}
\providecommand\BIBentryALTinterwordstretchfactor{4}
\providecommand\BIBentryALTinterwordspacing{\spaceskip=\fontdimen2\font plus
\BIBentryALTinterwordstretchfactor\fontdimen3\font minus
  \fontdimen4\font\relax}
\providecommand\BIBforeignlanguage[2]{{%
\expandafter\ifx\csname l@#1\endcsname\relax
\typeout{** WARNING: IEEEtran.bst: No hyphenation pattern has been}%
\typeout{** loaded for the language `#1'. Using the pattern for}%
\typeout{** the default language instead.}%
\else
\language=\csname l@#1\endcsname
\fi
#2}}

\bibitem{Donoho}
D.~L. Donoho, ``Compressed sensing,'' \emph{{IEEE} Trans. Inf. Theory},
  vol.~52, no.~4, pp. 1289--1306, 2006.

\bibitem{Tipping_RVM}
M.~E. Tipping, ``The relevance vector machine,'' in \emph{Advances in Neural
  Information Processing Systems}, vol.~12, 2000.

\bibitem{BCS}
S.~Ji, Y.~Xue, and L.~Carin, ``Bayesian compressive sensing,'' \emph{{IEEE}
  Trans. Signal Process.}, vol.~56, no.~6, pp. 2346--2356, 2008.

\bibitem{wipfRao_MMV}
D.~Wipf and B.~Rao, ``An empirical {B}ayesian strategy for solving the
  simultaneous sparse approximation problem,'' \emph{{IEEE} Trans. Signal
  Process.}, vol.~55, no.~7, pp. 3704--3716, 2007.

\bibitem{prasadjointSISO}
R.~Prasad, C.~R. Murthy, and B.~D. Rao, ``Joint approximately sparse channel
  estimation and data detection in {OFDM} systems using sparse bayesian
  learning,'' \emph{{IEEE} Trans. Signal Process.}, vol.~62, no.~14, pp.
  3591--3603, 2014.

\bibitem{parkerVolcanSchniter}
J.~T. Parker, V.~Cevher, and P.~Schniter, ``Compressive sensing under matrix
  uncertainties: An approximate message passing approach,'' in \emph{2011
  Conference Record of the Forty Fifth Asilomar Conference on Signals, Systems
  and Computers (ASILOMAR)}.\hskip 1em plus 0.5em minus 0.4em\relax IEEE, 2011,
  pp. 804--808.

\bibitem{prasadjointMIMO}
R.~Prasad, C.~R. Murthy, and B.~D. Rao, ``Joint channel estimation and data
  detection in {MIMO-OFDM} systems: A sparse bayesian learning approach,''
  \emph{{IEEE} Trans. Signal Process.}, vol.~63, no.~20, pp. 5369--5382, 2015.

\bibitem{sadeghidictionary}
M.~Sadeghi, M.~Babaie-Zadeh, and C.~Jutten, ``Dictionary learning for sparse
  representation: A novel approach,'' \emph{IEEE Signal Processing Letters},
  vol.~20, no.~12, pp. 1195--1198, 2013.

\bibitem{Wainwright}
A.~Pananjady, M.~J. Wainwright, and T.~A. Courtade, ``Linear regression with an
  unknown permutation: Statistical and computational limits,'' \emph{arXiv
  preprint arXiv:1608.02902}, 2016.

\bibitem{Unnikrish}
J.~Unnikrishnan, S.~Haghighatshoar, and M.~Vetterli, ``Unlabeled sensing with
  random linear measurements,'' \emph{arXiv preprint arXiv:1512.00115}, 2015.

\bibitem{Gibronv}
V.~Emiya, A.~Bonnefoy, L.~Daudet, and R.~Gribonval, ``Compressed sensing with
  unknown sensor permutation,'' in \emph{2014 IEEE International Conference on
  Acoustics, Speech and Signal Processing (ICASSP)}.\hskip 1em plus 0.5em minus
  0.4em\relax IEEE, 2014, pp. 1040--1044.

\bibitem{Caire2017}
S.~Haghighatshoar and G.~Caire, ``Signal recovery from unlabeled samples,''
  \emph{arXiv preprint arXiv:1701.08701}, 2017.

\bibitem{wieruchD2D}
D.~Wieruch, P.~Jung, T.~Wirth, and A.~Dekorsy, ``Determining user specific
  spectrum usage via sparse channel characteristics,'' in \emph{2015 49th
  Asilomar Conference on Signals, Systems and Computers}.\hskip 1em plus 0.5em
  minus 0.4em\relax IEEE, 2015, pp. 155--159.

\bibitem{tillmanncomputational}
A.~M. Tillmann, ``On the computational intractability of exact and approximate
  dictionary learning,'' \emph{IEEE Signal Processing Letters}, vol.~22, no.~1,
  pp. 45--49, 2015.

\bibitem{abid2017linreg}
A.~Abid, A.~Poon, and J.~Zou, ``Linear regression with shuffled labels,''
  \emph{arXiv preprint arXiv:1705.01342}, 2017.

\bibitem{wang2017}
G.~Wang, J.~Zhu, R.~S. Blum, P.~Braca, and Z.~Xu, ``Maximum likelihood signal
  amplitude estimation based on permuted blocks of differently binary quantized
  observations of a signal in noise,'' \emph{arXiv preprint arXiv:1706.01174},
  2017.

\bibitem{Wipf_Rao}
D.~P. Wipf and B.~D. Rao, ``Sparse {B}ayesian learning for basis selection,''
  \emph{{IEEE} Trans. Signal Process.}, vol.~52, no.~8, pp. 2153--2164, 2004.

\bibitem{hardypolyainequalities}
G.~H. Hardy, J.~E. Littlewood, and G.~P{\'o}lya, \emph{Inequalities}.\hskip 1em
  plus 0.5em minus 0.4em\relax Cambridge university press, 1952.

\bibitem{blumesmathDavies}
M.~Yaghoobi, T.~Blumensath, and M.~E. Davies, ``Dictionary learning for sparse
  approximations with the majorization method,'' \emph{IEEE Transactions on
  Signal Processing}, vol.~57, no.~6, pp. 2178--2191, 2009.

\bibitem{Ghahramani_mainlearning}
Z.~Ghahramani and G.~E. Hinton, ``\emph{Parameter estimation for linear
  dynamical systems},'' Tech. Rep., 1996.

\bibitem{mclachlan_EMalgo}
G.~McLachlan and T.~Krishnan, \emph{The {EM} algorithm and extensions}.\hskip
  1em plus 0.5em minus 0.4em\relax Wiley New York, 1997, vol. 274.

\bibitem{prasadbayesianGlbm}
R.~Prasad and C.~R. Murthy, ``Bayesian learning for joint sparse {OFDM} channel
  estimation and data detection,'' in \emph{Proc.\ Globecom}.\hskip 1em plus
  0.5em minus 0.4em\relax IEEE, 2010, pp. 1--6.

\bibitem{marques2009subspace}
M.~Marques, M.~Sto{\v{s}}i{\'c}, and J.~Costeira, ``Subspace matching: Unique
  solution to point matching with geometric constraints,'' in \emph{Computer
  Vision, 2009 IEEE 12th International Conference on}.\hskip 1em plus 0.5em
  minus 0.4em\relax IEEE, 2009, pp. 1288--1294.

\end{thebibliography}
\end{document}